\documentclass[fleqn,usenatbib]{mnras}

\usepackage{newtxtext,newtxmath}

\usepackage[T1]{fontenc}

\DeclareRobustCommand{\VAN}[3]{#2}
\let\VANthebibliography\thebibliography
\def\thebibliography{\DeclareRobustCommand{\VAN}[3]{##3}\VANthebibliography}

\usepackage{graphicx}	
\usepackage{amsmath}	

\newcommand{\Ex}{\mathcal{E}}                
\newcommand{\Exdot}{\dot{\mathcal{E}}}       

\newcommand{\Exabs}{\mathcal{E}_{\rm abs}}           
\newcommand{\Exdotarea}{\dot{\mathcal{E}}_{\rm area}}

\newcommand{\etaex}{\eta_{\mathrm{ex}}}              

\title[Photosynthetic exergy I]{Photosynthetic exergy I. Thermodynamic limits for habitable-zone planets}

\author[G. Covone \& A. Balbi]{
Giovanni Covone,$^{1,2,3}$\thanks{E-mail: giovanni.covone@unina.it}
and Amedeo Balbi,$^{4}$
\\
$^{1}$Dipartimento di Fisica "Ettore Pancini", Università di Napoli Federico II, Napoli, Italy\\
$^{2}$INFN, Sezione di Napoli, Complesso Universitario di Monte S. Angelo, Via Cintia Edificio 6, 80126 Napoli, Italy\\
$^{3}$INAF -- Osservatorio Astronomico di Capodimonte, via Moiariello 16, 80131 Napoli, Italy\\
$^{4}$Dipartimento di Fisica, Università di Roma Tor Vergata, Roma, Italy
}

\date{}

\pubyear{\the\year{}}

\begin{document}
\label{firstpage}
\pagerange{\pageref{firstpage}--\pageref{lastpage}}
\maketitle

\begin{abstract}
Photosynthesis is central to Earth's biosphere and a prime candidate for sustaining complex life on habitable exoplanets, yet a thermodynamically consistent treatment of the work potential of stellar radiation at planetary surfaces is still lacking. We develop a radiative–thermodynamic framework that quantifies the maximum useful work (exergy) extractable for a given star–planet configuration and yields exergy-based bounds on photosynthetic power and long-wavelength absorption cutoffs. From these we derive kinetically constrained red limits for high-$\Delta G$ photochemistry and apply them to Earth-like planets receiving the same bolometric flux from FGK and M blackbody hosts, computing thresholded photon supplies and truncated exergy fluxes below a photosystem II red limit. For such planets the constraints confine single-photon oxygenic photosynthesis to near-infrared bands around Solar-type stars and to somewhat bluer wavelengths around late M dwarfs. Integrated over the stellar spectrum, the thresholded photon supply and truncated exergy available to drive a photosystem water-oxidation step are larger by factors $\sim 5$ around FGK hosts than around $T_\star\approx 3000$~K M dwarfs. For the Solar–Earth system, the exergy-based upper bound on O$_2$ production exceeds the observed O$_2$ throughput by several orders of magnitude, consistent with Earth's photosynthetic efficiencies. Cool M dwarfs suffer a double penalty: fewer photons above threshold and a lower shortwave exergy fraction, yielding systematically tighter ceilings on high-$\Delta G$ photosynthesis than around FGK stars. Our framework provides upper limits on photosynthetically harvestable power on habitable-zone planets and enables comparisons of photosynthetic potential across exoplanetary systems, and can be extended to multi-band photosystems.
\end{abstract}



\section{Introduction}

The study of photosynthesis on exoplanets is a rapidly evolving field attracting increasing attention within the scientific community. 
This interest is two-fold: on one hand, the analysis of photoautotrophic potential provides a theoretical framework for quantifying the environmental conditions conducive to the evolution of complex life 
\citep[e.g.,][]{Gale2017, Cockell2016review, Lingam2019, Scharf2019}; on the other, photosynthetic processes represent a  source of potential remotely detectable biosignatures in exoplanetary atmospheres \citep[e.g.,][]{ Seager2005AsBio,Kiang_2007_AsBio...7..222K, Kiang_2018AsBio,Takizawa2017, Schwieterman2018AsBio}.

Analysing whether a given star–planet–atmosphere system can support sustained photosynthetic energy capture provides a physically grounded way to assess the environmental limits for surface life and the prospects for complex ecosystems beyond Earth \citep[e.g.,][]{Gale2017,Lingam2019,Covone2021}.

Terrestrial photosynthetic organisms generally operate by harnessing  photosynthetically active radiation (PAR)
in the $400 \lesssim \lambda \lesssim 700$ nm interval, with a flux $\simeq 4 \times 10^{20}$ photons m$^{-2}$ s$^{-1}$.
A likely minimum photon flux for sustained O$_2$-evolving
photosynthesis is $\sim 0.1\,\mu\mathrm{mol\,m^{-2}\,s^{-1}}$,
with an absolute lower bound of order
$0.01\,\mu\mathrm{mol\,m^{-2}\,s^{-1}}$
\citep{Littler1986,Cockell2009}.
These values are often adopted as a working hypothesis to investigate the feasibility of photosynthesis around other stars \citep{Lehmer2018, Lingam2019, Gale2017, Covone2021,Takizawa2017}. 
However, it is unclear whether the features of terrestrial pigments are universal \citep{Wolstencroft2002, Kiang_2007_AsBio...7..222K, Lingam2021}. 
Furthermore, photosynthesis metrics based solely on photon counts overlook the quality of photons, that is the fraction of radiant energy that can be converted into useful work. This fraction is quantified by the thermodynamic function exergy. A spectrum–resolved exergy treatment provides a natural way to express this work potential as a function of wavelength \citep[e.g.,][]{Petela_1964,Petela2003,Petela2008,Chen2008} and to set principled limits that allow stellar and atmospheric contexts to be compared on an equal
footing. In the context of exoplanets, \citet{Scharf2019} and \citet{Covone2021} were the first to apply radiative exergy to the efficiency of photosynthesis.

The present work builds upon this approach, with the aim of addressing some shortcomings of previous analyses.
First, the PAR theoretical limits are almost
always analysed in terms of energy or photon number rather than
exergy, so they do not distinguish between high- and low-quality
photons \citep[see, e.g.,][]{Lingam2019}. 
Second, requirements on reaction speed and free-energy margin are
rarely coupled explicitly to stellar spectral constraints: 
most studies either optimise pigment absorption at fixed energetic margins, or quantify
the spare free-energy required for redox and kinetic constraints
without including realistic stellar or atmospheric spectra
\citep[e.g.,][]{Marosvolgyi2010,Mielke2013,Lehmer2021FrASS,Duffy2023MNRAS}.
Third, there is still no simple framework that connects stellar spectral energy distributions (SED), atmospheric transmission, and pigment
absorption bands into quantitative upper bounds on high-activation (i.e., large-$\Delta G$) photochemical rates and efficiencies. 
This last gap is especially important for cool M-dwarf hosts, which, at fixed bolometric flux, may suffer a ``double penalty'': their redshifted spectra provide both fewer above-threshold photons and a smaller fraction of low-entropy, high-exergy radiation shortward of the photosynthetic red edge.

In this paper we develop an exergy-based framework that links the stellar spectrum, atmospheric filtering, and absorption band gap into a single radiative–thermodynamic budget. 
Here we explore the general thermodynamic bounds; in a companion paper, we will apply the framework to realistic stellar spectra, atmospheres and pigment action spectra.

In particular, we focus on high-activation  photochemical
reactions, using water oxidation to O$_2$ in oxygenic photosynthesis
as a benchmark. Within this framework, the long-wavelength limit of
oxygenic photosynthesis is no longer a fixed biochemical constant,
but an emergent property of the star–planet–atmosphere environment.
This benchmark reaction is both the kinetic bottleneck of oxygenic
photosynthesis and the main source of atmospheric O$_2$ biosignatures.

The paper is organised as follows. In Section~2 we introduce the general radiative--exergy framework. In Section~3 we derive single-photon red limits for high-$\Delta G$ photochemistry.
Section~4 applies these results to Earth analogues at fixed bolometric flux. Section~5 discusses the astrobiological implications and finally Section~6 summarises our main conclusions.

\section{Methods}
\label{sec:methods}

We consider rocky, Earth-like planets on circular orbits at distance $d$ in the temperate zone of main-sequence FGKM host-stars, such that long-term surface liquid water is possible. The planetary surface is characterised by the environmental temperature $T_{\rm env}$.
At the top of the atmosphere the planet receives a stellar spectral flux $F_{\rm TOA,\lambda}$, while the atmosphere is characterised by a wavelength-dependent transmittance $\mathcal{T} (\lambda)$ (including Rayleigh scattering, molecular absorption, and, where relevant, aerosol and cloud contributions). 
The resulting surface spectral flux is then
\begin{equation}
  F_{\lambda}  \;=\;
  \mathcal{T} (\lambda)\,F_{\rm TOA,\lambda} 
  \, = \mathcal{T} (\lambda)\, \left(\frac{R_\star}{d}\right)^2 F_{\star,\lambda} \, ,
\end{equation}
where $R_\star$ is the stellar radius and $F_{\star,\lambda}$ is the hemispherically integrated flux  at the host-star surface.
Throughout this work we use orbit- and diurnally averaged fluxes. 

The exergy of a radiation field is the maximum useful work that can be
obtained when that field is brought reversibly into equilibrium with an environment at temperature $T_{\rm env}$. For a generic radiation field with spectral energy flux $F_\lambda(\lambda)$ and spectral entropy flux $S_\lambda(\lambda)$, the spectral exergy flux is
\citep[see, for instance,][]{Candau2003}
\begin{equation}
  \dot{\mathcal{E}}_\lambda(\lambda)
  \;=\;
  F_\lambda(\lambda) \;-\;
  T_{\rm env}\,S_\lambda(\lambda) \, - \left( F_\lambda^{\rm env}
 - T_{\rm env} \, S_{\lambda}^{\rm env}\right) \, ,
  \label{eq:Ex_spec_def}
\end{equation}
where $F_\lambda^{\rm env}$ and $S_{\lambda}^{\rm env}$ are the spectral energy and entropy fluxes of a blackbody field at the ambient temperature.
In our model, the relevant environment in the definition of exergy is the biosphere temperature. Hence, we adopt $T_{\rm env}=288$\,K.

We note that in the so-called shortwave approximation, the contribution of environmental radiation fields 
in the visible–NIR is negligible. Hence we neglect the upward shortwave radiative field from the surface–atmosphere system. 
At $T_{\rm env}\simeq 288\,$K the upwelling shortwave emission is negligible in $0.35$–$1.6\,\mu\mathrm{m}$;  neglecting it changes $\Exdot$ by less than about $0.1\%$.
Therefore, we use the simplified form
\begin{equation}
  \dot{\mathcal{E}}_\lambda(\lambda)
  \simeq
  F_\lambda(\lambda) - T_{\rm env} S_\lambda(\lambda).
\end{equation}
Hence, the total exergy flux of the radiation field is
\begin{equation}
  \dot{\mathcal{E}}
  \;=\;
  \int_0^\infty
  \dot{\mathcal{E}}_\lambda(\lambda)\,{\rm d}\lambda
  \;=\;
  \int_0^\infty
  \bigl[F_\lambda(\lambda) - T_{\rm env} S_\lambda(\lambda)\bigr]\,
  {\rm d}\lambda\,.
  \label{eq:Ex_total_def}
\end{equation}

It is useful to introduce the quantity $\etaex(\lambda)$, defined as the maximum fraction of a photon’s energy that can, in principle, be converted into useful work at the environmental temperature $T_{\rm env}$.
The corresponding maximum work per absorbed photon is
\begin{equation}
  W_{\max}(\lambda)
  \;=\;
  \etaex(\lambda)\,\frac{hc}{\lambda}\,.
  \label{eq:wmax}
\end{equation}
The quantity $\etaex(\lambda)$ depends on both the source spectrum and \(T_{\mathrm{env}}\).
Then, the spectral exergy flux density can be written as
\(\dot{\mathcal{E}}_\lambda = \eta_{\mathrm{ex}}(\lambda)\,F_\lambda\). 
In practice, we are interested in a broad wavelength range $B = [\lambda_1,\lambda_2]$ rather than at a single $\lambda$. It is then useful to define band-integrated energy and exergy fluxes as
\begin{equation}
  F_B \;\equiv\; \int_{\lambda_1}^{\lambda_2} F_\lambda\,{\rm d}\lambda \, ,
  \qquad
  \dot{\mathcal{E}}_B \;\equiv\;
  \int_{\lambda_1}^{\lambda_2} \dot{\mathcal{E}}_\lambda\,{\rm d}\lambda \, ,
\end{equation}
and the corresponding band-averaged exergy efficiency
\begin{equation}
  \eta_{\rm ex}^{(B)}
  \;\equiv\;
  \frac{\dot{\mathcal{E}}_B}{F_B}
  \;=\;
  \frac{\displaystyle \int_{\lambda_1}^{\lambda_2}
    \eta_{\rm ex}(\lambda)\,F_\lambda(\lambda)\,{\rm d}\lambda}
  {\displaystyle \int_{\lambda_1}^{\lambda_2}
    F_\lambda(\lambda)\,{\rm d}\lambda}.
  \label{eq:eta_band}
\end{equation}
For a near-blackbody radiation field whose brightness temperature is approximately constant across the band, $T_b(\lambda)\simeq T_b$, the
band-averaged efficiency is well approximated by the Petela-type factor \citep{Petela_1964}
\begin{equation}
  \eta_{\rm ex}^{(B)} \simeq
  1 - \frac{4}{3}\frac{T_{\rm env}}{T_b}
    + \frac{1}{3} \, \left(\frac{T_{\rm env}}{T_b}\right)^{4} \, .
  \label{eq:eta_band_petela}
\end{equation}
For instance, for the solar case in the PAR band, exergy and energy fluxes differ only by an almost constant multiplicative factor \citep{Covone2021}.

Any photosynthetic system at the surface is described here by an effective dimensionless absorptance $a(\lambda)$, which selects the wavelengths that are actually absorbed ($0 \le a(\lambda) \le 1$); in this paper we only use idealised windows.

In this paper we focus on the exergy of the radiation that is actually absorbed by an idealised photosynthetic system, rather than the total incident stellar exergy \citep[as done in ][]{Scharf2019} or the exergy restricted to a fixed PAR band \citep[as done in ][]{Covone2021}. The absorber-specific exergy is the quantity entering the bounds derived below.

\section{Exergy--corrected and kinetically constrained red limits}
\label{sec:redlimit}
 
In this section we derive general energetic, exergy--corrected and kinetically constrained red limits for high-$\Delta G$ photochemical transformations. These results will later be applied to Earth analogues orbiting FGKM hosts to quantify their impact on the available photon and exergy budgets.
We consider a photochemical reaction with a Gibbs free-energy change $\Delta G$ per event, realised in the ideal limit by absorbing a minimal number $N_q$ of photons, each with wavelength $\lambda$ and energy $hc/\lambda$. Throughout, $\Delta G$ and the kinetic overhead $A$ are expressed per event (e.g., per electron transferred or per molecule of stoichiometric product).

\subsection{Exergy and kinetic constraints on $\lambda_{\max}$}

As only a fraction of the radiant energy can, even in principle, be converted to useful work because the environment has a non-zero
temperature and the radiation field is not in thermodynamic equilibrium with it, the exergy-corrected feasibility condition for a photochemical step with Gibbs free-energy change $\Delta G$ per event, driven by $N_q$ photons of wavelength $\lambda$, is
\begin{equation}
  \etaex(\lambda)\,\frac{hc}{\lambda}
  \;\ge\;
  \frac{\Delta G}{N_q}\,.
  \label{eq:redlimit_ex}
\end{equation}

For an endergonic photochemical step, $\Delta G > 0$ is both the maximum chemical free-energy stored in the products and the minimum useful work that the absorbed photons must supply in the reversible limit.

A purely energetic accounting, assuming reversible thermodynamic conditions, instead demands that the total energy carried by $N_q$ photons be at least $\Delta G$.
If one neglects the fact that a fraction of the incident power is thermodynamically
unusable as work (i.e., ignores exergy losses), this gives the
familiar energetic red limit
\begin{equation}
  \lambda_{\max}^{(E)} = \frac{h \, c \,  N_{\rm q}}{\Delta G}.
  \label{eq:lambda_E_max}
\end{equation}
In this case every photon is assumed to be fully convertible into work, $\etaex(\lambda)\equiv 1$. However, only the useful part of the photon energy can drive chemical work without entropy production.

Real transformations operating at a target turnover rate $r$ generally require an additional free-energy margin $A$ beyond $\Delta G$ to overcome activation barriers and other irreversibilities (electrochemical overpotentials, losses in the exciton-transfer steps, charge--transfer inefficiencies).
We denote by $r$ the microscopic reaction rate (turnover rate) of the photochemical step, i.e., the number of reaction events per unit time per active centre (with units of s$^{-1}$).

The energy $A$ is an additional free energy term (a kinetic overpotential or thermodynamic affinity) that must be supplied on top of $\Delta G$ to sustain a target turnover rate $r$ rather than a near–equilibrium rate $r_0$.
A minimal, thermally motivated scaling for the required overhead is
\begin{equation}
  A \;\gtrsim\; k_{\rm B} \,  T_{\rm env} \, 
  \ln\!\left(\frac{r}{r_0}\right) \, ,
  \label{eq:A_arrhenius}
\end{equation}
where $r$ is the operating reaction rate and $r_0$ is a reference rate near equilibrium. 
This follows from standard transition–state theory and the associated Arrhenius description of reaction rates: if each event is driven by an additional free energy $A$, the rate scales as $r \sim r_0 \exp (A/k_{\rm B}T_{\rm env})$, which in turn implies the bound above (see, e.g., \citealt{Atkins2010}, Chapter~28). These estimates represent generous upper limits: real biological systems are expected to dissipate additional free energy through exciton losses, charge–transfer inefficiencies and metabolic overheads, and will therefore operate below these upper limits.

We do not fix a unique microscopic turnover rate $r$ in this work: our model applies to any given $r$, with the kinetic overpotential $A$ absorbing the dependence on how far the system operates from equilibrium. 
Typical biological rates nevertheless provide a useful sanity check. 
For photosystem~II, reaction–centre turnover frequencies of order $10^2$–$10^3$ s$^{-1}$ at saturating light are commonly reported, while near–equilibrium water oxidation proceeds many orders of magnitude more slowly. 
At $T_{\rm env} = 288$ K, an overpotential $A \sim 0.3$ eV corresponds to $r/r_0 \sim 10^{4}$–$10^{6}$ in Eq.~\eqref{eq:A_arrhenius}, which is fully compatible with such far–from–equilibrium accelerations.

In this sense, overpotentials in the range $A = 0$–$0.3$ eV per electron are both thermodynamically and biologically reasonable.
The single-photon feasibility condition then generalises to
\begin{equation}
  \eta_{\rm ex}(\lambda)\,\frac{hc}{\lambda}
  \;\ge\; \frac{\Delta G + A}{N_q} ,
  \label{eq:kinetic_ineq}
\end{equation}
with $A$ constrained by Eq.~\eqref{eq:A_arrhenius}. 
%
%
The condition $\eta_{\rm ex}(\lambda) hc/\lambda \ge (\Delta G + A)/N_q$ states that each absorbed photon must provide at least this amount of usable free energy (per electron) in order for the transformation to proceed at the chosen far-from-equilibrium rate.

This inequality naturally defines the $\lambda_{\max}$, the longest wavelength for which the kinetically constrained condition is satisfied.
Formally, $\lambda_{\max}$ is the solution of
\begin{equation}
  \eta_{\rm ex}(\lambda)\,\frac{hc}{\lambda}
  \;=\;
  \frac{\Delta G + A}{N_q}\,.
\end{equation}
This definition is implicit and must be evaluated numerically once the radiation field (and so $\eta_{\rm ex}(\lambda)$) is specified. 
For a fixed $r$, the effect is a blue–shift of the usable red edge and a reduction of 
$\lambda_{\max}$ relative to the reversible estimate.

We will refer to three related red limits  on the usable wavelength:
\begin{itemize}
  \item the purely energetic red limit with $\eta_{\rm ex}(\lambda)\equiv 1$ and
  $A=0$, giving the closed form $\lambda^{(E)}_{\max}$ in
  Eq.~(\ref{eq:lambda_E_max});
  \item the exergy-corrected red limit $\lambda_{\max}(0)$ with $A=0$;
  \item the kinetically constrained red limit $\lambda_{\max}(A)$ for a chosen kinetic overhead $A>0$. 
\end{itemize}

In what follows we interpret $A$ as a kinetic overhead, i.e. an extra free-energy margin required to drive the reaction away from equilibrium. Our fiducial value $A \simeq 0.3$ eV per electron is not arbitrary: the oxygen–evolving complex in photosystem~II operates with an overpotential of $\sim 0.3$ V relative to the reversible water–oxidation potential \citep[e.g.,][]{Hayashi2016}, and classical thermodynamic analyses of primary photosynthesis likewise find spare margins of a few tenths of an eV per charge–separation step to suppress back–reactions and maintain fast turnover \citep[e.g.,][]{Mauzerall1976}. 
In the examples below we consider two representative cases: a near–reversible limit with $A = 0$ and a strongly driven case with $A = 0.3$ eV per electron.

Our analysis  connects to previous analyses of the long-wavelength limit of oxygenic photosynthesis based on day--night stability rather than exergy.  
In a recent three-level model of chlorophyll excitation and product formation, \citet{vanGrondelleBoeker2017} assume that each absorbed photon rapidly relaxes to the lowest excited state, and that part of this energy is stored as chemical free energy. In their model both the storage efficiency and a simple day–night stability metric depend exponentially on the energetic margin between the absorbing level and the stored free energy. 
This margin plays the same role as our overpotential $A$: efficient and stable operation require an energy headroom $A \gg k_{\rm B}T$ per electron. When the primary band is shifted beyond $\sim 900$ nm under a solar spectrum, this
condition is no longer met, yielding a natural long-wavelength limit for oxygenic photosynthesis on Earth.
The exergy-based condition (\ref{eq:redlimit_ex})
makes this energetic margin explicit and extends the red-limit analysis
from the solar case and $N_q=1$ to arbitrary stellar spectra, photon multiplicities and planetary environments.


In most discussions of the oxygenic-photosynthesis red limit, the focus is placed on the energetics of the H$_2$O/O$_2$ couple and on the architecture of the photosystems. 
Here we adopt a different angle:
we treat the red limit as an emergent quantity that depends jointly on the stellar spectrum, the atmospheric transmission window, and the Gibbs free energy of the target redox reaction. Within this exergy framework, changing the host star or the atmospheric window shifts both the usable exergy budget and the range of wavelengths over which high-$\Delta G$ photochemistry can be sustained.
 
Note that throughout this section we approximate the stellar spectrum by a blackbody at $T_\star$, in order to obtain simple analytic expressions for the red limits.
We have verified that replacing blackbody spectra with more realistic, non-blackbody stellar spectra changes the absorbed exergy $\Exabs$ and the shortwave exergy fractions by only $\sim 1$--$3\%$ (smallest for FGK, largest for late-M hosts).
This justifies the use of blackbody-based estimates in our analysis of the red limits.

\subsection{Consequences and limiting cases}

Let us consider the physically relevant limiting cases. In the reversible limit, $A = 0$ and $\eta_{\rm ex}(\lambda) = 1$, 
the general condition reduces to the purely energetic bound and we recover the familiar red limit $\lambda_{\max}^{(E)}$ of Eq.~(\ref{eq:lambda_E_max}).
For the actual radiation field, where $\eta_{\rm ex}(\lambda) < 1$, the exergy–corrected red edge $\lambda_{\max}(0)$ (with $A = 0$) is necessarily shorter.

Temperature, photon requirement and atmospheric transmission jointly determine the location of $\lambda_{\max}$. Increasing $T_{\rm env}$ reduces $\eta_{\rm ex}(\lambda)$ and increases the kinetic penalty $A \propto T_{\rm env}$, both acting to tighten the bound and push $\lambda_{\max}$ to shorter wavelengths. 
Conversely, a larger $N_q$ (multi–photon schemes) relaxes the per–photon requirement so that, in a purely energetic view, $\lambda_{\max}$ would shift to longer wavelengths, although the exergy factor partly offsets this trend. 
Finally, even if Eq. \eqref{eq:kinetic_ineq} is satisfied, the absorption band must lie within atmospheric transmittance windows, otherwise the system becomes photon-limited rather than exergy-limited.

For fixed chemical requirement $\Delta G$ and photon number $N_q$, the purely energetic red limit $\lambda_{\max}^{(E)} = hcN_q/\Delta G$ does not depend on the host star.
This degeneracy is broken once we account for exergy and kinetics.
The exergy factor $\eta_{\rm ex}(\lambda)$ depends on the radiation field, so that cooler stars have a smaller shortwave exergy fraction.
The kinetically constrained condition
$\eta_{\rm ex}(\lambda) hc/\lambda \ge (\Delta G + A)/N_q$ further tightens the red edge for realistic far-from-equilibrium rates.

The arguments above concern per–photon constraints. When combined with the actual spectral energy distribution, they imply that around cool M dwarfs this spectral tightening is compounded by the shape of the stellar SED. 
Most of the flux emerges at $\lambda \gtrsim 1\ \mu{\rm m}$, so the thresholded photon supply $\dot{N}_{\gamma,{\rm eff}}$ and the sub-$\lambda_{\rm thr}$ exergy $\dot{\mathcal{E}}_{\rm area}(<\lambda_{\rm thr})$  both collapse once one imposes a physiologically motivated threshold $E_{\rm thr}$ (see Section 4).
In this sense, M-dwarf spectra face a double
penalty for high-activation photochemistry: a smaller useful work per photon (lower $\eta_{\rm ex}$) and a much smaller fraction of photons that even satisfy $hc/\lambda \gtrsim E_{\rm thr}$.

In summary, Eqs.~\eqref{eq:redlimit_ex} and \eqref{eq:kinetic_ineq} state that the longest usable wavelength for a given transformation is controlled by the useful per-photon work $\eta_{\rm ex}(\lambda) \, hc/\lambda$ and, if a far-from-equilibrium turnover rate is required, by an additional kinetic overhead $A$.

\subsection{Upper wavelength limit for photosynthesis}
\label{sec:upperlimit} 

Here we derive an upper bound on the long-wavelength limit of useful photons for oxygenic photosynthesis by requiring that the per-photon
exergy exceeds the Gibbs free energy required for water oxidation, under the assumption that a single absorbed photon drives primary
charge separation.
We  apply the above  constraints to photosynthesis on rocky planets orbiting main-sequence host stars.
The resulting inequalities limit the longest wavelength that can, even
in principle, sustain a transformation with free-energy change $\Delta G$ per photochemical event.

The physically allowed long-wavelength range for photosynthesis remains a debated topic, both empirically and theoretically
\citep[e.g.,][]{Kiang_2007_AsBio...7..222K,Marosvolgyi2010,vanGrondelleBoeker2017,Lehmer2021FrASS,Duffy2023MNRAS,Zhen2022}. On modern Earth, all known oxygenic phototrophs use reaction centres based on chlorophylls~\textit{a}, \textit{d} or \textit{f}, with primary charge separation in the red and far--red and oxygenic photosynthesis observed up to $\simeq 740$\,nm in far-red-acclimated cyanobacteria \citep{Hu1998,Behrendt2015,Mielke2011}.
At longer wavelengths (around 760--780\,nm) oxygen evolution can still be detected in some systems, but with strongly reduced quantum yield, and no oxygenic organism is known to rely on photons beyond about 740--750\,nm as its primary energy band \citep{Pettai2005, Kiang_2007_AsBio...7..222K}. 
By contrast, anoxygenic phototrophs that use bacteriochlorophylls can exploit much longer wavelengths: green sulfur and related bacteria exhibit in~vivo reaction-centre absorption bands out to $\sim 1015$--$1020$\,nm, and anoxygenic photosynthesis has been inferred at wavelengths approaching 1.0--1.02\,$\mu{\rm m}$
\citep{Trissl1993,Scheer2003,Kiang_2007_AsBio...7..222K}. 
Empirically, terrestrial phototrophs  approach but do not exceed $\sim 1.0~\mu{\rm m}$ as the long-wavelength limit for primary photochemistry, with no known pigment system that reliably supports primary charge–separation photochemistry beyond this range.

\begin{figure}
\centering
\includegraphics[width=\columnwidth]{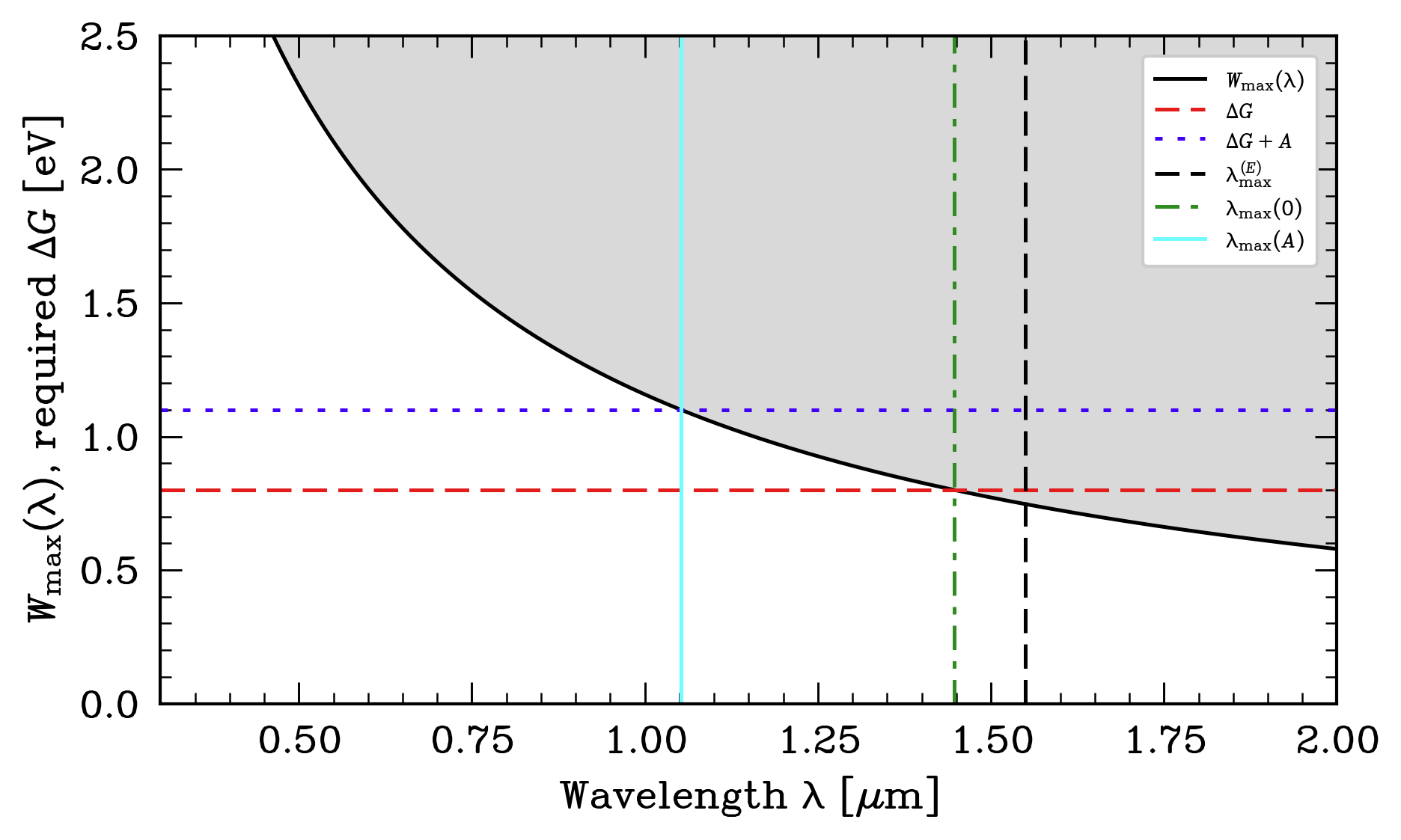}
\caption{Single-photon $\lambda$ limits for a PSII-like high-$\Delta G$ step around a Sun-like host at $T_{\rm env}=288$~K. The curve shows the maximum useful work per photon, $W_{\max}(\lambda)$; the shaded region marks where the required per-photon free energy exceeds $W_{\max}$ and is therefore thermodynamically forbidden. Horizontal lines indicate the benchmark requirement $\Delta G$ and the augmented requirement $\Delta G + A$;  vertical lines mark the corresponding $\lambda$ limits: the purely energetic limit 
$\lambda_{\max}^{(E)}$, the exergy-corrected    limit $\lambda_{\max}(0)$, and the exergy- and rate-constrained limit  $\lambda_{\max}(A)$.}
\label{fig:red_limit}
\end{figure}

As a benchmark for high-$\Delta G$ photochemistry we adopt a photosystem II (PSII)-like water-oxidation step with $\Delta G \simeq 0.8$\,eV per electron and a kinetic overhead $A \sim 0.3$\,eV per electron.
%
The value $\Delta G \simeq 0.8$\,eV is comparable to the free energy stored per absorbed photon in the oxygenic photosynthetic Z-scheme (the coupled PSII--PSI electron-transport chain) \citep[e.g.,][]{RossCalvin1967}, whereas $A \sim 0.3$\,eV is a typical kinetic overhead, corresponding to the overpotentials needed to sustain efficient water-oxidation catalysis in PSII and in analogous artificial systems \citep[e.g.,][]{OERreview2020}.

We use water oxidation as a benchmark high-activation photochemical step. This choice is motivated by three considerations.
First, the O$_2$/H$_2$O couple has a large Gibbs free-energy cost ($\Delta G_{\rm H_2O}\sim 3\times10^2 \, {\rm kJ}\, {\rm mol}^{-1}_{\mathrm{O_2}}$; see Appendix \ref{app:exergy_gibbs} for details), so that it probes the regime where exergy and kinetic penalties are most severe.
Second, water splitting is the bottleneck reaction of oxygenic photosynthesis and thus directly linked to the build-up of atmospheric O$_2$, the most widely discussed remotely detectable biosignature.
Third, both the equilibrium thermodynamics and the overpotentials of natural and artificial water-oxidation catalysts are well constrained, allowing the  parameters $\Delta G$ and $A$ in Section~3 to be anchored to experimentally measured values. The formalism itself, however, applies to any other high-activation photochemical step once $\Delta G$, $A$ and an appropriate threshold $E_{\rm thr}$ are specified.

Figure~\ref{fig:red_limit} provides a geometric illustration of these single-photon constraints for a Sun-like spectrum at
$T_{\rm env}=288$~K. The solid curve shows the maximum useful work per photon, $W_{\max}(\lambda) = \eta_{\rm ex}(\lambda)\,hc/\lambda$, while the horizontal lines at $\Delta G/N_q$ and $(\Delta G + A)/N_q$ represent, respectively, the purely energetic requirement and the energetic-plus-kinetic requirement for our PSII-like benchmark. 
Their intersections with $W_{\max}$ define the purely energetic red limit $\lambda_{\max}^{(E)}$, the exergy-corrected limit $\lambda_{\max}(0)$ and the exergy- and rate-constrained limit $\lambda_{\max}(A)$, which lies close to $\sim 1.0~\mu{\rm m}$ for the adopted parameters.

Independent molecular considerations provide a complementary benchmark. 
Efficient primary photochemistry requires that each absorbed photon be energetic enough to promote an electron across the pigment’s excitonic or charge–transfer gap (its effective electronic energy gap).

For typical conjugated systems, minimal gaps of order $\sim 1.1$\,eV are required for robust electronic excitation, implying a practical single-photon cutoff near $\lambda \approx 1.1~\mu{\rm m}$ \citep[e.g.,][]{Lingam2021}. 
This molecular estimate is broadly consistent with the thermodynamic red-edge values derived above and with the empirical absence of primary photochemistry beyond $\sim 1.0~\mu{\rm m}$ on Earth.

At $T_{\rm env} = 288$ K the shortwave (0.35–1.6 $\mu$m) exergy
flux is only $\sim 7\%$ lower than the corresponding shortwave
energy flux for solar-type stars, but up to $\sim 13\%$ lower for
cool M dwarfs, corresponding to $\eta_{\rm ex} \simeq 0.93$ and
$\eta_{\rm ex} \simeq 0.87$, respectively.
For a PSII-like water-oxidation step ($\Delta G\simeq 0.8$\,eV per electron, $N_q=1$), this shifts the single-photon exergy-corrected red edge from $\lambda_{\max,\odot}(0)\approx 1.44\,\mu{\rm m}$ to
$\lambda_{\max,{\rm M}}(0)\approx 1.35\,\mu{\rm m}$ and, for a realistic kinetic overhead $A\sim 0.3$\,eV, to $\lambda_{\max,\odot}(A)\sim 1.0\,\mu{\rm m}$ and $\lambda_{\max,{\rm M}}(A)\sim 0.95\,\mu{\rm m}$. In this sense, thermodynamics and kinetics alone already push the viable single-photon regime for high-$\Delta G$ photochemistry close to
$\sim 1.0~\mu{\rm m}$ around solar-type hosts, and slightly bluer for cool M dwarfs. 
 
Although the single–photon red limit itself is only mildly star dependent, the short-wavelength photon and exergy budgets above a given threshold are much poorer for cool M dwarfs than for solar-type hosts. For simple FGKM blackbodies scaled to the same bolometric flux, truncation at $\lambda_{\rm thr}\simeq 1.0~\mu{\rm m}$ leaves \(\simeq 70\%\) of the stellar energy flux for a Sun-like star but only \(\simeq 25\)–30\% for a $T_\star\simeq 3000$\,K M dwarf, and the corresponding shortwave photon flux is reduced by comparable factors (see discussion in next section). 
Once exergy weighting is included, the usable shortwave exergy around cool M dwarfs is suppressed by factors of a few relative to solar-type hosts, making it substantially harder to sustain high-$\Delta G$ photosynthesis at biologically relevant areal rates.

Multi-photon or multi-step architectures have been proposed to circumvent this constraint by partitioning the redox span into smaller increments or combining sub-gap photons. 
However, these strategies remain strongly constrained: each photon must have energy larger than the  pigment's effective electronic energy gap, to drive an electronic transition, while multi-photon schemes suffer from reduced effective cross-sections, more complex reaction networks, and increased kinetic losses.
Sequential or two-colour chains can extend the overall redox range only if each individual photosystem operates above its own excitonic gap. 

More exotic multi-photon pathways that would combine photons with energies much smaller than typical electronic excitation gaps remain highly speculative rather than demonstrated biophysical designs.
Proposed chained photosystems extending oxygenic photochemistry to $\sim 2.1\,\mu{\rm m}$ \citep[e.g.,][]{Lehmer2018} should therefore be regarded as extreme hypothetical scenarios rather than as realistic models based on known pigment systems.

Environmental filtering further compounds the quantum constraint. In Earth-like atmospheres, strong H$_2$O absorption bands beyond $\sim 1.1~\mu{\rm m}$ severely attenuate surface irradiance, further reducing the viability of long-wavelength photochemistry. Even if pigments could absorb at, say, 1300\,nm, the available photon and exergy flux at the surface would be strongly suppressed under many atmospheric conditions.

\section{Bounds on high-activation photochemistry}
\label{sec:high-activation}

In Section~3 we derive single-photon constraints on high-$\Delta G$ photochemistry, expressed as a maximum wavelength $\lambda_{\rm max}$ set by the Gibbs free-energy requirement $\Delta G$, the kinetic overpotential $A$, and the photon number $N_q$. 
In what follows we introduce a physiologically motivated activation threshold $\lambda_{\rm thr}$, which must satisfy $\lambda_{\rm thr} \leq \lambda_{\max}$.
Our goal in this section is to estimate how much high-activation chemistry can be driven at the planetary surface under these constraints.

We fold the single-photon limits together with the stellar spectral energy distribution $F_{\lambda}$ at the top of the atmosphere, the atmospheric transmittance $\mathcal{T}(\lambda)$, and an effective absorptance of the photosynthetic system $a(\lambda)$, together with a primary-step quantum yield $\Phi(\lambda)$ ($0 \le \Phi \le 1$). 
$\Phi(\lambda)$ denotes the fraction of absorbed photons that actually trigger the first useful photochemical event (e.g., primary charge separation in a reaction centre), rather than being lost to fluorescence or heat. 
Throughout this paper, pigments are treated in an idealised way: we assume generic absorbers that harvest all photons within a prescribed activation band, without specifying the particular molecular cross-section or the detailed photophysics. 
These bands act as placeholders for any high-activation photochemical pathway compatible with the given single-photon constraints.




We denote by $E_{\rm thr}$ the threshold photon energy for the photochemical step of interest (e.g. PSII primary charge separation) and by $\dot n_{\rm O_2}$  the areal production rate of molecular oxygen (mol\,m$^{-2}$\,s$^{-1}$). 

As above, we use a PSII-like water-oxidation step as our benchmark high-activation reaction. The expressions below, however, apply to any photochemical step once the required Gibbs free energy $\Delta G$, kinetic overpotential $A$ and
threshold photon energy $E_{\rm thr}$ are specified.
The Gibbs free energy per mole of O$_2$, evaluated as described in Appendix \ref{app:exergy_gibbs}, corresponds to $\Delta G \simeq 0.8~{\rm eV}$ per electron in the four-electron water-oxidation half-reaction.

The corresponding spectral photon number flux at the top of the atmosphere is \( F_{{\rm TOA},\lambda}\,\lambda/(hc)\), so that the above-threshold photon supply at the planetary surface, capable of driving a
reaction with threshold photon energy \(E_{\rm thr}\), is
\begin{equation}    
  \dot N_{\gamma,{\rm eff}} =
  \int_0^\infty F_{{\rm TOA},\lambda} \, 
  \frac{\lambda}{hc}\,a(\lambda) \,\mathcal{T}(\lambda)\,\Phi(\lambda)\,
  \Theta\bigl(hc/\lambda - E_{\rm thr}\bigr)\,{\rm d}\lambda \, .
\label{eq:N_eff}
\end{equation}
where $\Theta$ is the Heaviside step function, equal to 1 for $hc/\lambda \ge E_{\rm thr}$ and 0 otherwise.
This expression makes explicit the spectral ``quality filter'': even when the total photon count is large, the factor $\Theta(hc/\lambda - E_{\rm thr})$
removes all contributions with $\lambda > \lambda_{\rm thr}$. The quantum yield $\Phi(\lambda)$ ensures that only above-threshold photons that actually initiate the high-activation chemistry are counted.

We can now bound the areal throughput of high-activation chemistry, taking water oxidation as benchmark. The pigment-absorbed exergy per unit area that is in principle usable for such reaction is
\begin{equation}
\dot{\mathcal{E}}_{(< \lambda_{\rm thr}), \rm area} \;\equiv\;
\int_0^{\lambda_{\rm thr}}
\eta_{\rm ex}(\lambda) \, F_{\rm TOA,\lambda}\,\mathcal{T}(\lambda)\,a(\lambda)\;
{\rm d}\lambda \, ,
\label{eq:Earea_chem}
\end{equation}
since only photons with $\lambda \leq \lambda_{\rm thr}$  can provide enough free energy to drive the reaction. 
We refer to $\dot E^{(<\lambda_{\rm thr})}_{\rm area}$ as the truncated exergy flux per unit area.

\begin{figure}
\centering
\includegraphics[width=\columnwidth]{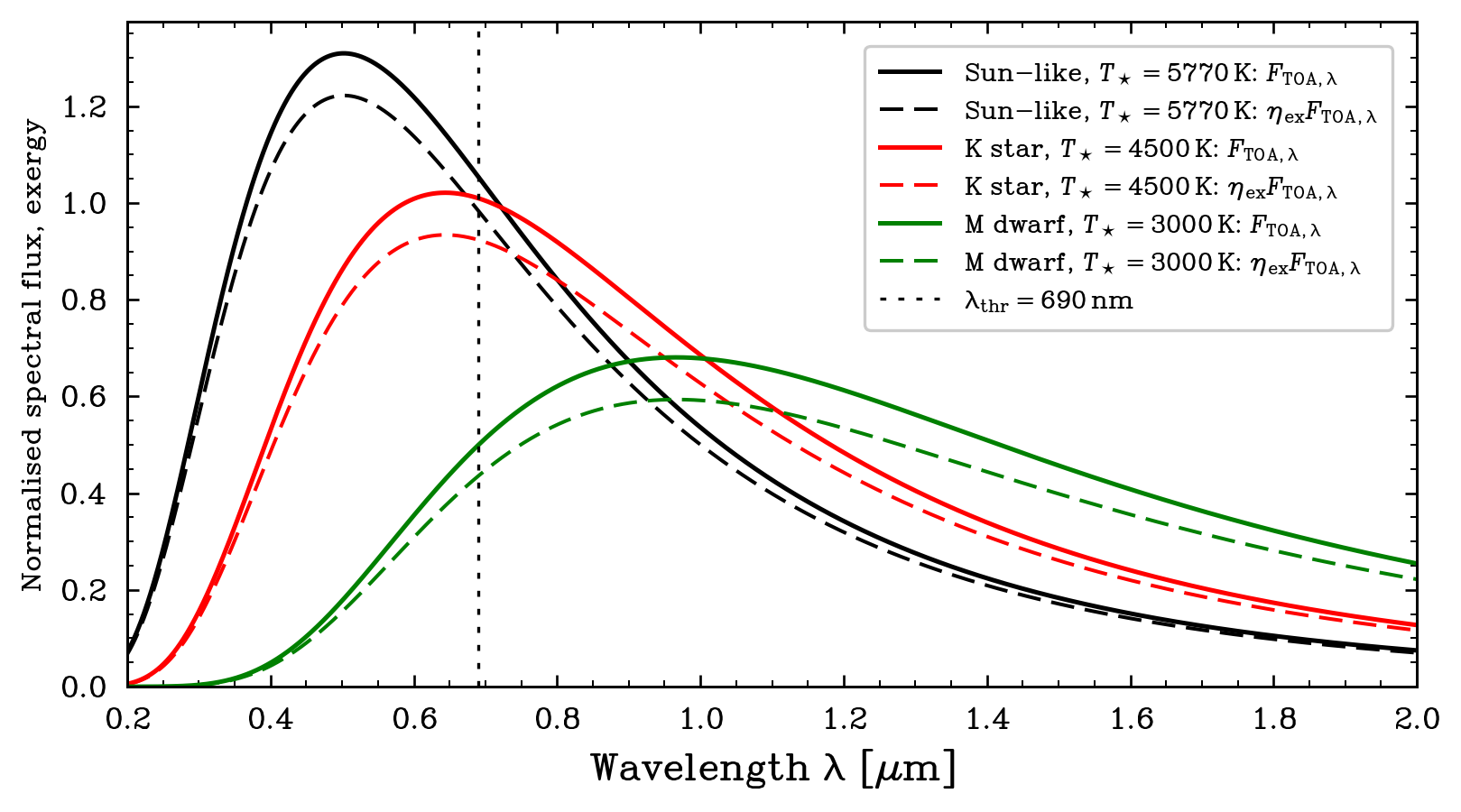}
\caption{Spectral flux and exergy at the top of the atmosphere for three blackbody models, all  normalized to the same bolometric flux $F_{\rm bol}=1361\,{\rm W\,m^{-2}}$.
Solid curves show $F_{\rm TOA,\lambda}$ and dashed curves $\eta_{\rm ex} F_{\rm TOA,\lambda}$.
The vertical line marks a PSII-like threshold at $\lambda_{\rm thr}=690$ nm, illustrating how the short-wavelength band available for high-$\Delta G$ photochemistry shrinks for M-type hosts.}
\label{fig:spectral_exergy}
\end{figure}

Figure~\ref{fig:spectral_exergy} illustrates how, at fixed $F_{\rm bol}$, the
top-of-atmosphere spectrum and its exergy-weighted counterpart shift with $T_\star$, and how the short-wavelength band below $\lambda_{\rm thr}=690$~nm shrinks towards cool M dwarfs. 
For this diagnostic plot we ignore atmospheric and pigment effects, setting
$\mathcal{T}(\lambda)=a(\lambda)=\Phi(\lambda)=1$, so that the curves reflect
only the stellar spectral shape and the exergy weighting.

Exergy weighting suppresses the long-wavelength tail and slightly blue-shifts and narrows the spectral peak for all stellar types. 
Quantitatively, for the three blackbody spectra in Fig.~\ref{fig:spectral_exergy} the exergy weighting shifts the peak wavelength blueward by
$\Delta\lambda_{\rm peak}\simeq 0.7$~nm ($T_\star=7000$~K), $1.1$~nm ($5770$~K) and $4.3$~nm ($3000$~K) for $T_{\rm env}=288$~K.
Therefore, for FGK stars the exergy-weighted curves remain close to the energy spectra, whereas the relative suppression of the near-infrared tail is largest for the 3000~K M-dwarf spectrum. 
Hence, at fixed $F_{\rm bol}$ the same total energy input is less useful in exergy terms for cooler hosts: visually, the area under the dashed curve, relative to the solid one, is smallest for the M dwarf. 
For FGK stars, $\lambda_{\rm thr}\simeq690$~nm falls within the bulk of both the energy and exergy spectra, leaving a broad band of photons above threshold, whereas for the M dwarf the threshold lies on the Wien side of the spectrum so that the usable exergy is confined to a much narrower short-wavelength shoulder. Since all curves are normalised to the same bolometric flux, these penalties are purely spectral-shape effects rather than differences in the total energy received.

Therefore, taken together, Eqs.~\eqref{eq:N_eff} and~\eqref{eq:Earea_chem} quantify the resulting double penalty faced by cool-star spectra: a collapse in the number flux of super-threshold photons and a collapse in the sub-$\lambda_{\rm thr}$ exergy that can be harvested even with ideal downstream conversion.

Let us consider two planets that both receive the same bolometric flux at the top of the atmosphere (e.g., $F_{\rm bol} \simeq 1361$~W\,m$^{-2}$), one orbiting a solar-type star ($T_\star = 5770$~K) and one orbiting a cool M dwarf with $T_\star = 3000$~K. For simplicity, assume $\mathcal{T}(\lambda)=1$, $a(\lambda)=1$ and $\Phi(\lambda)=1$  (i.e., no atmospheric absorption and a perfectly absorbing, perfectly efficient photosystem), and take a PSII-like activation threshold $E_{\rm thr}\simeq 1.8$~eV, corresponding to $\lambda_{\rm thr}\simeq 690$~nm. 
Under these assumptions, Eq.~\eqref{eq:N_eff} reduces to
\begin{equation}
\dot N_{\gamma,{\rm eff}} \;=\;
F_{\rm bol}\,\frac{R(T_\star)}{hc} \, ,    
\end{equation}
where
\begin{equation}
R(T_\star)\equiv
\frac{\displaystyle\int_0^{\lambda_{\rm thr}} B_\lambda(T_\star)\,\lambda\,{\rm d}\lambda}
     {\displaystyle\int_0^\infty B_\lambda(T_\star)\,{\rm d}\lambda} \, ,  
\end{equation}
and $B_\lambda$ is the Planck function. 
For a solar-type star we find $R(5770\,{\rm K})\simeq 2.4\times 10^{-7}$\,m, whereas for the M dwarf $R(3000\,{\rm K})\simeq 4.8\times 10^{-8}$\,m, so that $\dot N_{\gamma,{\rm eff}}$ is larger by a factor $\simeq 5$ in the solar case. 
Numerically, one obtains $ \dot N_{\gamma,{\rm eff}}^{\odot} \simeq 1.6\times 10^{21}\ {\rm photons\ m^{-2}\ s^{-1}}, $ and $\dot N_{\gamma,{\rm eff}}^{\rm M} \simeq 3.1\times 10^{20}\ {\rm photons\ m^{-2}\ s^{-1}}.$

The exergy fluxes show a similar contrast. If we approximate $\eta_{\rm ex}(\lambda)$ by a Petela-type factor that depends only on $T_\star$ and $T_{\rm env}=288$\,K, the shortwave exergy fraction is $\eta_{\rm ex}\simeq 0.93$ for a solar-type star and $\eta_{\rm ex}\simeq 0.87$ for a 3000\,K M dwarf. Combined with the fraction of the bolometric flux emitted at $\lambda<\lambda_{\rm thr}$ (about $0.48$ for the Sun and $0.08$ for the M dwarf), the truncated exergy fluxes are
\[
\Exdotarea^{(<\lambda_{\rm thr}),\odot}
\simeq 6.1\times 10^2\ {\rm W\ m^{-2}},
\qquad
\Exdotarea^{(<\lambda_{\rm thr}),M}
\simeq 9.8\times 10^1\ {\rm W\ m^{-2}},
\]
a ratio of $\simeq 6$ even under these idealised assumptions. 

\begin{figure}
\centering
\includegraphics[width=\columnwidth]{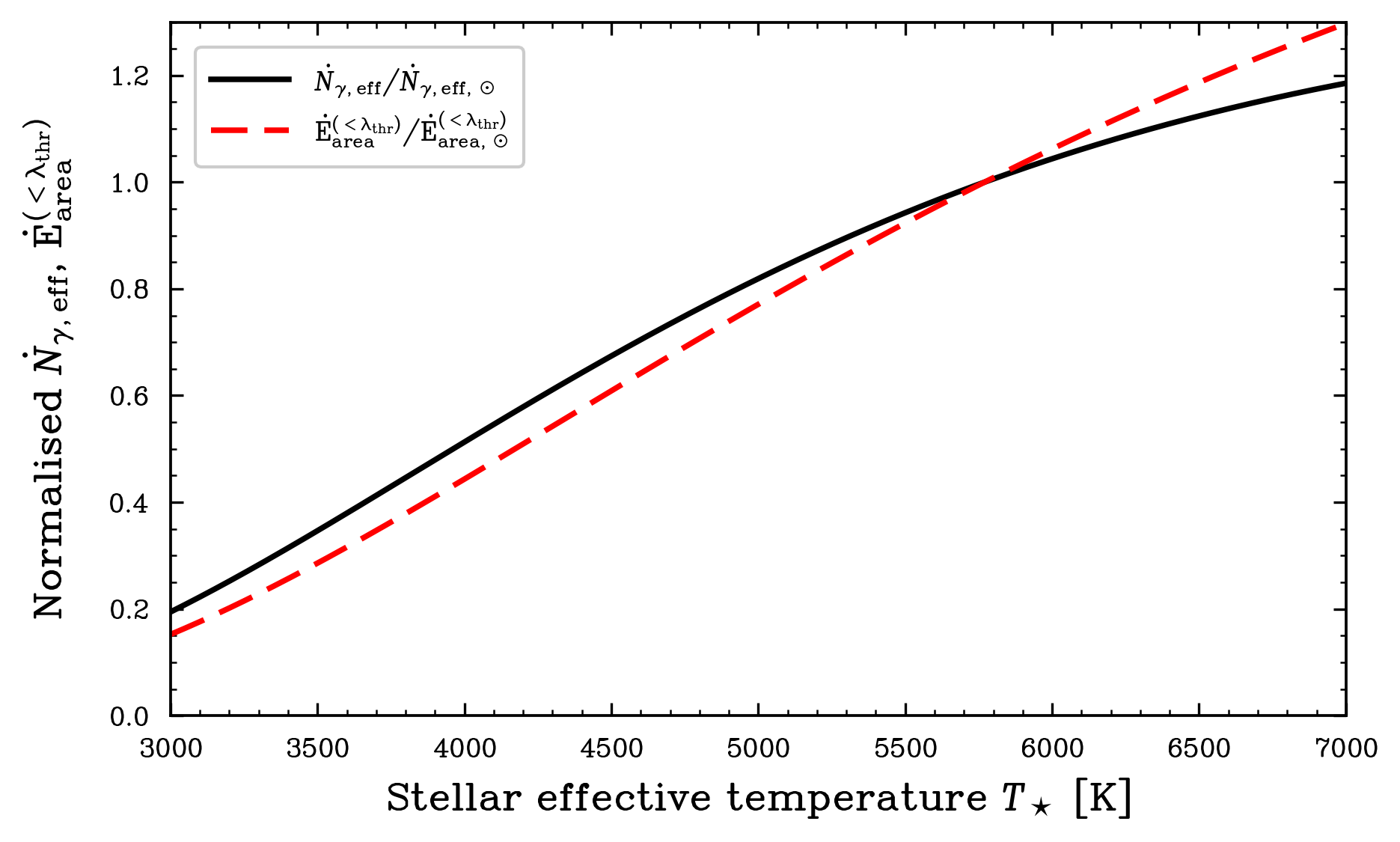}
\caption{Normalised photon and exergy upper limits for a high-$\Delta G$ reaction as a function of stellar effective temperature. We assume Earth analogues at fixed $F_{\rm bol}=1361\,{\rm W\,m^{-2}}$, $T_{\rm env}=288$ K and a PSII-like threshold $\lambda_{\rm thr}=690$ nm, with blackbody stellar spectra, no atmosphere, and only photons with $\lambda<\lambda_{\rm thr}$ included. The blue curve shows the thresholded photon rate $\dot N_{\gamma,{\rm eff}}/\dot N_{\gamma,{\rm eff},\odot}$; the orange curve the truncated exergy flux $\dot E^{(<\lambda_{\rm thr})}_{\rm area}/
 \dot E^{(<\lambda_{\rm thr})}_{\rm area,\odot}$;
both normalised to a Sun-like host.}
\label{fig:exergy_Tstar}
\end{figure}

To generalise this example, we evaluate the thresholded photon supply $\dot N_{\gamma,\rm eff}$ and the truncated exergy flux $\dot{\Ex}_{(<\lambda_{\rm thr}),\rm area}$ for blackbody FGKM host-stars, all scaled to the same bolometric flux. Figure~\ref{fig:exergy_Tstar} shows the resulting upper limits for high-$\Delta G$ photochemistry as a function of stellar effective temperature, normalised to a solar analogue.
 
At fixed $F_{\rm bol}$, both the thresholded photon supply and the truncated exergy increase modestly towards hotter F-type hosts (up to $\simeq 30$--$40\%$ above the solar case at $T_\star \simeq 7000\,\mathrm{K}$), and decline steeply towards late M dwarfs (down to $\sim 0.2$ of the solar $\dot N_{\gamma,{\rm eff}}$ and $\sim 0.15$ of the solar exergy upper limit at $T_\star = 3000\,\mathrm{K}$; see Fig.~\ref{fig:exergy_Tstar}). 
%
%
For a Solar spectrum in the 400--700 nm band we find that exergy- and photon-based measures of photosynthetically useful flux differ by only $\simeq 7\%$ (see Appendix~\ref{app:par_benchmark}), which explains why previous PAR photon-count metrics provide a good first approximation for Solar-type hosts.

\subsection{Exergy-limited oxygen production across stellar types}
\label{sec:exergy_O2_rates}

We can now derive explicit upper limits for the areal production rate of molecular oxygen. For a given star and bolometric flux, 
the usable exergy flux $\dot{\Ex}_{(<\lambda_{\rm thr}),{\rm area}}$ given by Eq.~(\ref{eq:Earea_chem}), 
carried by photons above the activation threshold can be related to the steady production of O$_2$ via water splitting, which must  satisfy the exergy work balance 
\begin{equation}
  \dot{\Ex}_{(<\lambda_{\rm thr}),{\rm area}}
  \;\ge\;
  \dot{n}_{\rm O_2}\,\Delta G_{\rm H_2O} \, ,
\end{equation}
which implies the upper bound
\begin{equation}
  \dot{n}^{\max}_{\rm O_2}
  \;=\;
  \frac{\dot{\Ex}_{(<\lambda_{\rm thr}),{\rm area}}}
       {\Delta G_{\rm H_2O}} \, .
  \label{eq:nO2max_def}
\end{equation}
%
This is an upper bound, as it ignores all irreversibilities downstream of photon capture (non-radiative losses, charge-transfer inefficiencies, metabolic overheads).

Inserting these values into Eq.~\eqref{eq:nO2max_def}, 
and using the value of $\Delta G_{\rm H_2O}$ defined above, yields
\begin{equation}
\begin{aligned}
  \dot{n}^{\max}_{\rm O_2}(<\lambda_{\rm thr})_\odot
  &\simeq 2.0\times 10^{-3}\,\mathrm{mol\,m^{-2}\,s^{-1}},\\
  \dot{n}^{\max}_{\rm O_2}(<\lambda_{\rm thr})_M
  &\simeq 3.0\times 10^{-4}\,\mathrm{mol\,m^{-2}\,s^{-1}}.
\end{aligned}
\label{eq:nO2_ceilings}
\end{equation}
so that an Earth analogue around a cool M dwarf can, at best, sustain only \(\sim 1/6\) of the maximal oxygenic photosynthetic throughput per unit area achievable around a solar-type host,
for the same incident \(F_{\rm bol}\).

This factor can be understood analytically from the spectral shape of the stellar emission. As shown in Appendix~\ref{app:wien_tail}, once the activation threshold $\lambda_{\rm thr}$ lies on the Wien side of the spectrum the
fraction $f_{<}(T_\star,\lambda_{\rm thr})$ of the bolometric flux emitted at $\lambda < \lambda_{\rm thr}$ scales approximately as
\begin{equation}
  f_{<}(T_\star,\lambda_{\rm thr}) \;\propto\;
  \exp\!\left[-\,\frac{hc}{\lambda_{\rm thr} k_{\rm B} T_\star}\right],
  \label{eq:f_less_asympt_main}
\end{equation}
so that the truncated exergy flux $\dot{\mathcal{E}}_{(<\lambda_{\rm thr}),{\rm area}}$ inherits an
exponential sensitivity to $T_\star$ at fixed $F_{\rm bol}$.  For
$\lambda_{\rm thr}=690\,\mathrm{nm}$ one has $x_{\rm thr}\simeq 3.6$
for the Sun and $x_{\rm thr}\simeq 7$ for a $3000$~K M dwarf
(Appendix~\ref{app:wien_tail}), which already implies a reduction
in the truncated flux, and hence in $\dot n^{\max}_{\rm O_2}$, by a factor of order six between the two cases, in good agreement with the numerical values in Eq.~\eqref{eq:nO2_ceilings}.

 
It is useful to compare these upper limits to Earth's present-day gross O$_2$ production. 
Taking a global primary productivity of $\sim 120$~PgC~yr$^{-1}$, the corresponding  O$_2$ production rate is $\sim 10^{16}$~mol~yr$^{-1}$, or
\begin{equation}
\langle \dot n_{\rm O_2}\rangle_{\oplus}
\simeq 6\times10^{-7}\ {\rm mol\,m^{-2}\,s^{-1}},
\end{equation}
when averaged over the Earth's surface area. 
Our Solar--Earth exergy-limited maximum  is therefore 
\begin{equation}
\frac{\dot n_{\rm O_2}^{\max}(<\lambda_{\rm thr})_{\odot}}
     {\langle \dot n_{\rm O_2}\rangle_{\oplus}}
\sim 3\times 10^{3}
\end{equation}
that is, about three orders of magnitude larger than the present-day mean total O$_2$ flux. In other words, Earth operates about three orders of magnitude below the thermodynamically allowed maximum set by the Solar irradiance and water-oxidation chemistry.

Detailed exergy budgets of modern plants provide a useful
reality check on these upper limits. For a representative
temperate crop species under non-stress conditions,
\citet{Silva2015ExergyPhotosynthesis} find an exergy
efficiency of $\simeq 12\%$ at the chloroplast level and $\simeq 4\%$ for the whole plant, with most of the exergy loss arising already at the light-harvesting and primary electron-transport stages. Even in the favourable Solar--Earth
case, only a few per cent of the incident PAR exergy can therefore be channelled into net chemical exergy of biomass and O$_2$ on planetary scales, so the values of $\dot n^{\max}_{\rm O_2}$ derived here should be read as strict upper envelopes rather than predictions for any particular biosphere.
%
%
%
More realistic assumptions about $T(\lambda)$, $a(\lambda)$ and $\Phi(\lambda)$ (e.g., strong H$_2$O bands and NIR-tuned pigments around M dwarfs) only increase this contrast.

%
%

The bounds derived above can be read biochemically as a statement about how often reaction centres can successfully perform high-activation steps such as water oxidation. 
For a fixed bolometric flux, the key control parameters are (i) the fraction of the
stellar spectrum that lies above the activation threshold $\lambda_{\rm thr}$ and (ii) the exergy efficiency $\eta_{\rm ex}(\lambda)$ of those photons. 
Around solar-type stars a substantial fraction of the flux is emitted at $\lambda<\lambda_{\rm thr}$ and with high $\eta_{\rm ex}$, so both the thresholded photon supply $\dot N_{\gamma,{\rm eff}}$ and the truncated exergy $\Exdotarea^{(<\lambda_{\rm thr})}$ are large. 
Around cool M dwarfs, most of the flux is instead emitted at $\lambda\gg\lambda_{\rm thr}$, with energy less than the 
activation threshold estimated above and contributes negligibly to 
$ \dot N_{\gamma,{\rm eff}} $
and  $\dot{\mathcal{E}}^{(<\lambda_{\rm thr})}_{\rm area}.$
The bottleneck is therefore the \emph{quality} of the photons, not the total photon count.
 
The dependence on the underlying chemistry enters only through $\Delta G$ and through the choice of $E_{\rm thr}$ and the kinetic overpotential $A$. 
For phototrophic metabolisms that exploit lower-potential redox couples (e.g., typical anoxygenic photosynthesis), one has smaller $\Delta G$ and often a somewhat lower $E_{\rm thr}$, so that the same formalism yields larger values of $\dot n^{\max}$ for a given radiative environment. 
However, the qualitative trends remain unchanged: cool stars still suffer from a sharp reduction in $\dot{N}_{\gamma,{\rm eff}}$ and 
$\dot{\mathcal{E}}_{(<\lambda_{\rm thr}),{\rm area}}$ whenever the activation threshold sits in the visible, and any high-$\Delta G$ step will be more strongly
penalised around M dwarfs than around FGK stars.



\section{Discussion and astrobiological implications}
\label{sec:astro_implications}

A central message of this work is that the red wavelength limit of oxygenic photosynthesis is not set solely by the chemistry of the H$_2$O/O$_2$ couple, but by the entire star–planet environment. In our framework the long-wavelength cutoff reflects both the spectral exergy supply from the irradiating star and the Gibbs free-energy requirement and operating temperature of the target reaction.



\subsection{Oxygenic photosynthesis and O$_2$ biosignatures}
\label{subsec:oxy_biosignatures}

For oxygenic photosynthesis driven by a PSII-like water-oxidation step, taken together, thermodynamic, kinetic, and quantum constraints confine the viable primary absorption bands to a relatively narrow window.
The above red-limit inequalities, applied to $\Delta G \simeq 0.8$~eV per electron and a biologically motivated overpotential $A\sim 0.3$~eV, place the exergy- and rate-aware single-photon cutoff near 
$\lambda_{\max}\sim 1.0~\mu$m for a solar-type star and somewhat bluer for cool M dwarfs. 
In the near-reversible case ($A=0$) the corresponding difference is $\Delta\lambda_{\max}\simeq 0.09\,\mu{\rm m}$.
Independently, the requirement that each absorbed photon promote an electron across the excitonic or
charge–transfer gap of a realistic pigment favours minimal gaps of order $\sim 1.1$~eV, implying a practical single-photon cutoff near
$\lambda\sim 1.0$--$1.1~\mu$m. 

Empirical constraints from terrestrial phototrophs are broadly consistent with these theoretical long-wavelength limits.
Oxygenic photosynthesis on Earth relies on chlorophylls with
primary absorption bands in the red and far-red, and no known
oxygenic organism uses photons beyond $\sim$750\,nm as its main energy source \citep[e.g.,][]{Kiang_2007_AsBio...7..222K,Mielke2011}.
Anoxygenic phototrophs employing bacteriochlorophylls can extend primary absorption into the near-infrared, with in vivo reaction-centre bands detected up to $\sim$1015--1020\,nm and
anoxygenic photosynthesis inferred at wavelengths approaching $1.0$--$1.02\,\mu$m \citep{Kiang_2007_AsBio...7..222K}.
These empirical cutoffs are consistent with our exergy- and rate-constrained single-photon limits near $\lambda_{\max}\sim 1.0$--$1.1\,\mu$m for FGK hosts (Section~3.3), and somewhat bluer for late M dwarfs, supporting the plausibility of the exergy-based bounds derived here.

Together with the empirical absence of primary photochemistry beyond $\sim 1.0~\mu$m in terrestrial phototrophs, this argues that oxygenic photosystems cannot simply shift their primary absorption bands arbitrarily deep into the near-infrared. 
In particular, speculative oxygenic edges at $\lambda\gtrsim 1.2$--$1.5~\mu$m would likely require either non-standard multi-photon architectures or very small electronic gaps, and are not naturally accommodated within the single-photon exergy framework developed here.

In Section~\ref{sec:high-activation} we further showed that even when these per-photon constraints are satisfied, high-activation chemistry such as water  oxidation faces a stringent supply-side limitation around cool stars. At fixed bolometric flux, the thresholded photon supply $\dot N_{\gamma,\mathrm{eff}}$ and the truncated exergy $\dot E^{(<\lambda_{\rm thr})}_{\rm area}$ available to drive water oxidation are smaller around a $\sim 3000$ K M dwarf by a factor of order five compared to a Sun-like star, implying a similar reduction in the exergy-limited O$_2$ throughput per unit area. Our solar benchmark $\dot n^{\max}_{\rm O_2}(<\lambda_{\rm thr})_\odot \simeq 2\times 10^{-3}\,{\rm mol\,m^{-2}\,s^{-1}}$ already lies several orders of magnitude above standard estimates of Earth's gross O$_2$ production, so these bounds should be read as generous upper limits rather than predictions. 
Real biospheres are expected to operate at only a small fraction of this exergy ceiling, yet the factor-of-few suppression around M dwarfs is already apparent at the level of these optimistic upper bounds (see Fig.~\ref{fig:exergy_Tstar}).



Taken together, our results imply that, for oxygenic photosynthesis:
(i) the primary absorption bands, and thus any oxygenic ``red edge'' biosignature, are likely confined to $\lambda\lesssim 1.0$--$1.1~\mu$m even on planets orbiting cool stars; 
and (ii) for a fixed bolometric stellar flux and similar surface conditions, the shortwave exergy available to drive water oxidation is systematically larger around FGK stars than around late M dwarfs. 
This does not exclude oxygenic biospheres around M dwarfs, but it does suggest that achieving and maintaining Earth-like atmospheric O$_2$ levels may require more favourable conditions (e.g., higher photosynthetic productivity per unit area, larger areal coverage of phototrophs, or more efficient O$_2$ retention) than in the FGK case. Our exergy-based bounds provide a physically motivated prior on which star–planet systems are most promising for strong oxygenic biosignatures.

\subsection{Other phototrophic pathways and general exergy metrics}
\label{subsec:other_pathways}

The formalism developed here is not restricted to water oxidation. The only
inputs from the underlying chemistry are the Gibbs cost per event $\Delta G$,
the kinetic overpotential $A$, and the appropriate activation threshold
$E_{\rm thr}$ for the primary photochemical step. 
For phototrophic metabolisms that exploit lower-potential redox couples (e.g.\ many forms of anoxygenic photosynthesis), the net photochemical step typically involves a smaller $\Delta G$ and can often operate with a somewhat lower $E_{\rm thr}$ than PSII-like oxygenic pathways.

Inserting these values into Eqs.~\eqref{eq:N_eff}--\eqref{eq:Earea_chem} and \eqref{eq:nO2max_def} yields larger values of $\dot n^{\max}$ for a given radiative environment: in other words, low-$\Delta G$ photochemistry is thermodynamically easier to support, and its exergy ceiling lies further above any plausible biological throughput.

However, the qualitative trends with stellar type remain unchanged. For any high-activation step whose threshold lies in the visible or short near-IR, cool M dwarfs still suffer from a sharp reduction in the thresholded photon supply $\dot N_{\gamma,{\rm eff}}$ and in the truncated exergy $\dot E_{\rm area}^{(<\lambda_{\rm thr})}$ relative to FGK stars at the same bolometric flux. 
The ``double penalty'' identified in Section~\ref{sec:high-activation} (that is, fewer photons above the threshold and a smaller exergy fraction per photon) applies to any phototrophic metabolism with comparable $E_{\rm thr}$, even if its net redox span $\Delta G$ is smaller than that of water oxidation.

These results suggest a family of exergy-based metrics for planetary habitability and biosignatures. One can define, for instance, a
high-activation photochemical capacity per unit area,
$\dot{\mathcal{E}}_{\rm area}(<\lambda_{\rm thr})$, or the corresponding
exergy-limited throughput $\dot n^{\max}$ for a specific redox couple, and
compare these across star–planet combinations. Such quantities are more
informative than raw photon counts because they incorporate both the spectral
distribution of the stellar flux and the thermodynamic difficulty of the
target reaction. In the companion paper we apply this framework to
realistic pigment action spectra and atmospheric models, constructing synthetic biosignature predictions (e.g., red-edge locations and maximum O$_2$ levels) across a grid of stellar types and planetary conditions.

\section{Conclusions}
\label{sec:conclusions}

In this paper we have derived new thermodynamic constraints on photochemical reactions in irradiated planetary environments. 
At the single-photon level, the inequalities Eqs.~\eqref{eq:redlimit_ex}--\eqref{eq:kinetic_ineq} show how the Gibbs free-energy change $\Delta G$, kinetic overpotential $A$ and photon multiplicity $N_q$ together fix a maximum usable wavelength for a given photochemical step. 
At the area-averaged
level, Eqs.~\eqref{eq:N_eff}--\eqref{eq:nO2max_def} translate these per-photon limits into bounds on the thresholded photon
supply, the truncated exergy flux and the maximum throughput of
high-activation reactions such as water oxidation. Taken together, these results make explicit that the long-wavelength limit of oxygenic photosynthesis is not a fixed biochemical constant, but an emergent property of the coupled star–planet–atmosphere system.
%

These bounds are strict upper limits set by radiative exergy and activation thresholds; they do not attempt to model specific biospheres. 
In particular, the Solar--Earth benchmark shows that present-day Earth operates about three orders of magnitude below the exergy-based
ceiling for O$_2$ production; 
these bounds should therefore be interpreted as generous thermodynamic upper limits rather than as
predictions for any specific biosphere.
Within this framework, cool M dwarfs suffer a double penalty for high-$\Delta G$ photochemistry: fewer photons above the relevant thresholds and a lower shortwave exergy fraction, implying systematically tighter ceilings on oxygenic photosynthesis and other demanding redox pathways than around FGK stars at the same bolometric flux.

In this sense, our exergy-based red limit adds an “energy-availability” parameter to multidimensional habitability frameworks in which stellar composition and bioessential elements already play a central role \citep[e.g.,][]{Hinkel2020ApJ,Covone2025metals}.

In real pigment--reaction-centre complexes the exergy- and kinetically constrained red limits derived here must still be compatible with the microscopic electronic structure: an
absorbed photon has to promote an electron across the relevant excitonic or charge-transfer gap. 
The formalism itself is general and can be applied to any photochemical step by specifying $\Delta G$, $A$ and $E_{\rm thr}$, yielding exergy-based metrics of high-activation photochemical capacity per unit area that are more informative than photon counts alone for exoplanet comparisons. 
In a companion paper we will combine these bounds with realistic stellar spectra, atmospheric transmittance and pigment action spectra to obtain quantitative estimates of photosynthetic fluxes and associated biosignatures across a grid of star--planet systems.

\section*{Acknowledgements}
The authors thank the anonymus referee for the careful reading.  
This work was supported by the University of Napoli Federico II (project: FRA-CosmoHab, CUP E65F22000050001).

\section*{Data Availability}
No new data were generated or analysed in this study.


\bibliographystyle{mnras}
\bibliography{biblio_exobiology}




\appendix

\section{Gibbs cost of water oxidation}
\label{app:exergy_gibbs}

For the benchmark high-activation step we adopt the water-oxidation
half–reaction
\begin{equation}
  \mathrm{O_2} + 4\mathrm{H}^+ + 4\mathrm{e}^- \rightarrow 2\mathrm{H_2O} .
  \label{eq:O2_half_reaction_app}
\end{equation}
The corresponding reduction potential at $T = 298\,$K, as a function
of pH and oxygen partial pressure $p_{\rm O_2}$, is given by the
Nernst equation
\begin{equation}
  E(\mathrm{pH}, p_{\rm O_2}) =
  E^\circ(298\,\mathrm{K})
  - 0.05916\,\mathrm{V}\,\mathrm{pH}
  + \frac{0.05916\,\mathrm{V}}{4}\,
    \log_{10}\!\left(\frac{p_{\rm O_2}}{\mathrm{bar}}\right),
  \label{eq:nernst_O2_app}
\end{equation}
with $E^\circ(298\,\mathrm{K}) = 1.229\,$V at $\mathrm{pH}=0$ and
$p_{\rm O_2} = 1\,$bar.  The Gibbs free energy per mole of
$\mathrm{O_2}$ required for water oxidation is
\begin{equation}
  \Delta G_{\mathrm{H_2O}}(\mathrm{pH}, p_{\rm O_2})
  = n F E = (4F)\,E
  \approx (385.94 \,\mathrm{kJ\,V^{-1}\,mol^{-1}})\,E ,
  \label{eq:deltaG_O2_app}
\end{equation}
where $F = 96485\,\mathrm{C\,mol^{-1}}$ is Faraday's constant.

In this paper we are interested in Earth-like surface conditions. Evaluating Eq.~\eqref{eq:nernst_O2_app} at $T = 298\,\mathrm{K}$,
$\mathrm{pH} = 7$ and $p_{\rm O_2} = 0.21\,$bar gives
$E \simeq 0.805\,\mathrm{V}$ and hence
\begin{equation}
  \Delta G_{\mathrm{H_2O}} \simeq
  3.11 \times 10^{2}\,\mathrm{kJ\,mol^{-1}_{\mathrm{O_2}}}
  \quad (\simeq 0.80\,\mathrm{eV\ per\ electron}),
  \label{eq:deltaG_O2_value_app}
\end{equation}
which is the value adopted in Section~4.
For moderate departures from these reference conditions, the
dependence on pH is well approximated by a linear shift
$\Delta G_{\mathrm{H_2O}} \propto E \propto
E^\circ - 0.05916\,\mathrm{V}\,\mathrm{pH}$,
so that $\Delta G_{\mathrm{H_2O}}$ decreases by
$\simeq 23\,\mathrm{kJ\,mol^{-1}}$ per unit increase in pH at 298 K.
Corrections due to non-ideal activities are small compared to the astrophysical uncertainties considered in this work and are neglected here.

\section{Near-proportionality and Solar--Earth PAR benchmark}
\label{app:par_benchmark}

For any absorbed radiation field, the photon rate $\dot N_\gamma$, 
the absorbed radiant power $F_{\rm abs}$ and the band-averaged exergy 
efficiency $\eta_{\rm ex}^{(B)}$ are related by
\begin{equation}
  \dot{\mathcal{E}}_{\rm abs}
  = \eta_{\rm ex}^{(B)} F_{\rm abs}
  = \eta_{\rm ex}^{(B)} \,\langle E_{\rm ph} \rangle \,\dot N_\gamma,
  \label{eq:par_scaling}
\end{equation}
where $\dot{\mathcal{E}}_{\rm abs}$ is the absorbed exergy flux and 
$\langle E_{\rm ph} \rangle$ is the mean absorbed photon energy in the  band of interest. Over the wavelength range relevant for photosynthesis 
these quantities vary only weakly, so exergy-, energy- and photon-based measures of ``useful'' flux are nearly proportional.

As a concrete benchmark, consider the present Solar--Earth system in the 400--700 nm (PAR) band, with a top-hat absorptance 
$a(\lambda) = 1$ in this interval and $a(\lambda) = 0$ elsewhere, 
and $T(\lambda) = 1$. Modelling the Sun as a blackbody at 
$T_\star = 5770\,\mathrm{K}$ and normalising to 
$F_{\rm bol} = 1361\,\mathrm{W\,m^{-2}}$, the absorbed radiant power 
per unit area in 400--700 nm is
\begin{equation}
  E_{\rm abs}^{\rm (PAR)} =
  \int_{400\,\mathrm{nm}}^{700\,\mathrm{nm}}
  F_{\rm TOA,\lambda}\,{\rm d}\lambda
  \simeq 5.0 \times 10^{2}\,\mathrm{W\,m^{-2}},
  \label{eq:par_Eabs}
\end{equation}
about $0.37$ of the bolometric flux.
The corresponding photon rate is
\begin{equation}
  \dot N_{\gamma}^{\rm (PAR)} =
  \int_{400\,\mathrm{nm}}^{700\,\mathrm{nm}}
  F_{\rm TOA,\lambda}\,
  \frac{\lambda}{hc}\,{\rm d}\lambda
  \simeq 1.4 \times 10^{21}\,\mathrm{m^{-2}\,s^{-1}},
  \label{eq:par_Ngamma}
\end{equation}
about $0.23$ of the total photon flux, implying a mean photon energy
\begin{equation}
  \langle E_\gamma \rangle_{\rm PAR} \equiv
  \frac{E_{\rm abs}^{\rm (PAR)}}{\dot N_{\gamma}^{\rm (PAR)}}
  \simeq 2.3\,\mathrm{eV}.
  \label{eq:par_meanE}
\end{equation}

For a blackbody source at $T_\star$ and an environment at $T_{\rm env}$, the band-averaged exergy efficiency is well approximated by the Petela factor (Eq.~8), which for $T_\star = 5770\,\mathrm{K}$ and $T_{\rm env} = 288\,\mathrm{K}$ gives $\eta_{\rm ex}^{\rm (bb)} \simeq 0.93$. 
Approximating the PAR-band $\eta_{\rm ex}(\lambda)$ by this value yields an absorbed exergy flux
\begin{equation}
  \dot{\mathcal{E}}_{\rm abs}^{\rm (PAR)}
  \simeq 0.93\, E_{\rm abs}^{\rm (PAR)}
  \simeq 4.7 \times 10^{2}\,\mathrm{W\,m^{-2}}.
  \label{eq:par_exergy}
\end{equation}

Thus, in the Solar--Earth PAR band the exergy-based metric differs from the raw absorbed radiant power by only $\sim 7\%$, and both are nearly proportional to the photon rate $\dot N_{\gamma}^{\rm (PAR)}$. For Solar-like irradiation in the  visible, exergy- and photon-based measures of photosynthetically  useful flux are therefore almost interchangeable.

\section{Wien-tail approximation for the truncated flux}
\label{app:wien_tail}

For a blackbody host-star with temperature $T_\star$ and bolometric flux $F_{\rm bol}$ at the top of the atmosphere, the fraction of the radiant energy carried by wavelengths shorter than a threshold $\lambda_{\rm thr}$ is
\begin{equation}
  f_{<}(T_\star,\lambda_{\rm thr}) \equiv
  \frac{1}{F_{\rm bol}}
  \int_{0}^{\lambda_{\rm thr}} B_\lambda(T_\star)\,{\rm d}\lambda,
  \label{eq:f_less_def_app}
\end{equation}
with $B_\lambda$ the Planck function.  Introducing
\begin{equation}
  x_{\rm thr} \equiv
  \frac{hc}{\lambda_{\rm thr} k_{\rm B} T_\star},
  \label{eq:x_thr_def_app}
\end{equation}
and using the Wien approximation $B_\lambda \propto \lambda^{-5}
\exp[-hc/(\lambda k_{\rm B}T_\star)]$ valid when $x_{\rm thr}\gtrsim 3$,
a change of variable $y = hc/(\lambda k_{\rm B} T_\star)$ yields
\begin{equation}
  f_{<}(T_\star,\lambda_{\rm thr})
  \simeq
  \frac{15}{\pi^{4}} \int_{x_{\rm thr}}^{\infty} y^{3} e^{-y}\,{\rm d}y
  =
  \frac{15}{\pi^{4}} e^{-x_{\rm thr}}
  \bigl(x_{\rm thr}^{3} + 3x_{\rm thr}^{2}
        + 6x_{\rm thr} + 6\bigr).
  \label{eq:f_less_full_app}
\end{equation}
For $x_{\rm thr}\gg 1$ this reduces to the simple asymptotic form
\begin{equation}
  f_{<}(T_\star,\lambda_{\rm thr})
  \sim
  \frac{15}{\pi^{4}}\,x_{\rm thr}^{3} e^{-x_{\rm thr}},
  \label{eq:f_less_asympt_app}
\end{equation}
which underlies the exponential scaling quoted in Eq.~\eqref{eq:f_less_asympt_main}.

\bsp	
\label{lastpage}
\end{document}